\pdfoutput=1

\documentclass[prl,aps,twocolumn,showpacs,preprintnumbers,amssymb,nofootinbib,superscriptaddress,10pt]{revtex4-1}
\usepackage{hyperref,graphicx,array,multirow,color,amsmath}

\hypersetup{pdftitle={},pdfcreator={},linkcolor=[rgb]{0.15,0.35,0.75},colorlinks=true,citecolor=[rgb]{0.675,0,0.2},urlcolor=[rgb]{0.15,0.35,0.65}}

\newcommand{\fwbox}[2]{\text{\makebox[#1][c]{$\hspace{-150pt}\displaystyle#2\hspace{-150pt}$}}}
\newcommand{\fwboxL}[2]{\text{\makebox[#1][l]{$#2$}}}
\newcommand{\fwboxR}[2]{\text{\makebox[#1][r]{$#2$}}}
\newcommand{\bigger}[1]{\raisebox{-0.95pt}{\scalebox{1.25}{$#1$}}}
\renewcommand{\phi}{\varphi}

\newcommand{\eq}[1]{\vspace{-3.5pt}\begin{equation}\hspace{-0pt}#1\hspace{-0pt}\vspace{-3.5pt}\end{equation}}
\newcommand{\eqL}[1]{\eq{\fwboxL{0pt}{\hspace{-115pt}#1}}}
\newcommand{\fig}[3]{\raisebox{#1}{\includegraphics[scale=#2]{#3}}}
\DeclareMathOperator*{\Res}{\mathrm{Res}}
\DeclareMathOperator*{\Ord}{\mathrm{Ord}}
\newcommand{\x}[2]{(#1,#2)}
\renewcommand{\u}[2]{(\hspace{-0.5pt}#1;\hspace{-1.5pt}#2\hspace{-0.5pt})}
\newcommand{\mi}{\raisebox{0.75pt}{\scalebox{0.75}{$\hspace{-2pt}\,-\,\hspace{-0.5pt}$}}}
\renewcommand{\pl}{\raisebox{0.75pt}{\scalebox{0.75}{$\hspace{-2pt}\,+\,\hspace{-0.5pt}$}}}
\newcommand{\weier}{Weierstra\ss~}
\newcommand{\proj}[1]{\left[#1\right]}
\newcommand{\traintrack}{\mathfrak{T}^{(L)}}
\definecolor{hblue}{rgb}{0,0,0.575}
\definecolor{hred}{rgb}{0.575,0.0,0.225}

\begin{document}
\title{\texorpdfstring{Traintracks Through Calabi-Yaus: Amplitudes Beyond Elliptic Polylogarithms \\[-18pt]~}{Traintracks Through Calabi-Yaus: Amplitudes Beyond Elliptic Polylogarithms}}
\author{Jacob~L.~Bourjaily}
\affiliation{Niels Bohr International Academy and Discovery Center, Niels Bohr Institute,\\University of Copenhagen, Blegdamsvej 17, DK-2100, Copenhagen \O, Denmark}
\author{Yang-Hui He}
\affiliation{School of Physics, NanKai University, Tianjin, 300071, P.R.~China}
\affiliation{Department of Mathematics, City, University of London, EC1V 0HB, UK}
\affiliation{Merton College, University of Oxford, OX14JD, UK}
\author{Andrew~J.~McLeod}
\affiliation{Niels Bohr International Academy and Discovery Center, Niels Bohr Institute,\\University of Copenhagen, Blegdamsvej 17, DK-2100, Copenhagen \O, Denmark}
\author{Matt~von~Hippel}
\affiliation{Niels Bohr International Academy and Discovery Center, Niels Bohr Institute,\\University of Copenhagen, Blegdamsvej 17, DK-2100, Copenhagen \O, Denmark}
\author{Matthias~Wilhelm}
\affiliation{Niels Bohr International Academy and Discovery Center, Niels Bohr Institute,\\University of Copenhagen, Blegdamsvej 17, DK-2100, Copenhagen \O, Denmark}

\begin{abstract}
We describe a family of finite, four-dimensional, $L$-loop Feynman integrals that involve weight-$(L\pl1)$ hyperlogarithms integrated over $(L\mi1)$-dimensional elliptically fibered varieties we conjecture to be Calabi-Yau. At three loops, we identify the relevant K3 explicitly; and we provide strong evidence that the four-loop integral involves a Calabi-Yau threefold. These integrals are necessary for the representation of amplitudes in many theories---from massless \mbox{$\varphi^4$ theory} to integrable theories including maximally supersymmetric Yang-Mills theory in the planar limit---a fact we demonstrate.
\end{abstract}
\maketitle

\section{Introduction}\label{introduction_section}\vspace{-14pt}

The study of scattering amplitudes is on some level the study of classes of special functions. This is true even at tree level where, although tree-level scattering amplitudes are generally rational, the analytic, geometric, and combinatoric aspects of these functions have become a rich source of insight (see e.g.\ ref.\ \cite{ArkaniHamed:2012nw}). Beyond leading order, loop integration generally results in transcendental functions, which have been the subject of extensive research in recent years.

The simplest of these `new' functions are polylogarithms and their generalization to `hyperlogarithms' \cite{goncharov:hyperlogs}. Our understanding of such functions has grown enormously in recent years due to a rich interplay between number theory, analysis, and algebraic geometry (see e.g.\ refs.\ \cite{Brown:2009qja,Brown:2009ta,Goncharov:2009kx}). This has fueled corresponding advances in physics, and today many of the most impressive reaches into perturbation theory are predicated on  an explicit or implicit assumption about the polylogarithmic nature of certain classes of integrals \cite{Arkani-Hamed:2014via,Caron-Huot:2016owq,Dixon:2016apl,Chicherin:2017dob,Caron-Huot:2018dsv}.

Next in complexity are iterated integrals involving elliptic curves, often referred to as elliptic polylogarithms. These integrals have begun to yield to systematic analysis, and many of the tools previously exclusive to hyperlogarithms, such as coactions and symbols, have been generalized to the elliptic case \cite{brown2011multiple,Adams:2017ejb,Remiddi:2017har,Broedel:2017kkb,Broedel:2018iwv}.

In this Letter, we describe a class of Feynman integrals in massless $\phi^4$ theory that seem to saturate the potential complexity required by (virtually any) four-dimensional quantum field theory. Specifically, we study the following (conventionally normalized) $L$-loop Feynman integral involving $(2L\pl6)$ massless fields:
\vspace{-6pt}\eq{\fwbox{0pt}{\fwboxR{0pt}{\traintrack\equiv\!\!}\fig{-22.6pt}{1}{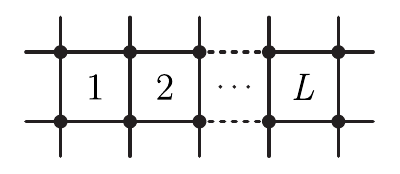}.}\label{traintrack_intro}\vspace{-6pt}}
We call these integrals `traintracks' due to their obvious resemblance. At one loop, this is the famous `four-mass box' integral first evaluated in \mbox{ref.\ \cite{Hodges:boxInt}}; at two loops, it is the elliptic double-box integral studied in \mbox{ref.\ \cite{Bourjaily:2017bsb}}. For higher loops, these traintrack integrals involve irreducible components defined on higher-dimensional algebraic varieties. Thus, these traintracks lay out a path of increasing complexity: from polylogarithms, to elliptic polylogarithms, to K3 surfaces and so on. This is not the first time that such complexities have been seen in the study of scattering amplitudes. Indeed, a similar sequence (also Calabi-Yau) has been observed for massive, two-dimensional sunset integrals \cite{Bloch:2014qca,Bloch:2016izu,broadhurstprivate}. The traintrack (\ref{traintrack_intro}) represents the simplest instance of this complexity in the context of massless theories in four dimensions.

Although defined in the context of massless $\phi^4$ theory, the relevance of these integrals to a wider class of quantum field theories is immediate from the point of view of generalized unitarity \cite{Bern:1994cg,Bern:1994zx} (see also \mbox{refs.\ \cite{Bourjaily:2015jna,Bourjaily:2017wjl}}), in which (\ref{traintrack_intro}) represents an independent loop integrand relevant to many processes in general four-dimensional theories. 
Because it can appear as a sub-topology of higher-loop, lower-multiplicity integrals, it is clearly relevant to even the simplest processes (see e.g.\ refs.\ \cite{Bourjaily:2015bpz,Bourjaily:2016evz}).

Of perhaps more interest to some readers, it turns out that the traintrack integral (\ref{traintrack_intro}) represents the {\it entire} leading-order contribution to a component amplitude of planar maximally supersymmetric Yang-Mills ($\mathcal{N}\!=\!4$ SYM) theory. We prove this fact using the relationship between amplitudes in planar $\mathcal{N}\!=\!4$ SYM theory and its deformation to an integrable fishnet theory, within which (\ref{traintrack_intro}) arises more directly.  It is safe to say that traintracks have some relevance to amplitudes in virtually all quantum field theories in four dimensions at sufficiently high loop order or multiplicity.

\vspace{-10pt}\section{Feynman Parameterization}\vspace{-10pt}

Let us first derive a Feynman-parametric representation for the integral (\ref{traintrack_intro}) that makes manifest both its weight $(2L)$ and its dual-conformal invariance. To do this, we first express the $a$th particle's momentum as the difference $p_a\!\equiv\!(x_{a+1}\mi x_a)$ between `dual-momentum' $x$-coordinates (with cyclic labeling understood). We also associate the $i$th loop momentum with the dual point $x_{\ell_i}$. In terms of these, we define 
\eq{\x{a}{b}\equiv(x_a\mi x_a)^2\quad\text{and}\quad\x{\ell_i}{a}\equiv(x_{\ell_i}\mi x_a)^2.}
These dual points are associated with the (Poincar\'e) dual of the Feynman graph, which we will label by
\eq{\hspace{-20pt}\fwbox{0pt}{\fig{-22.6pt}{1}{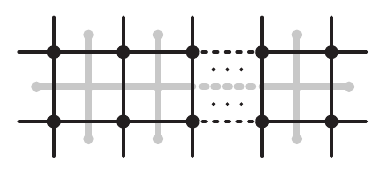}\!\!\bigger{\Leftrightarrow}\!\fig{-22.6pt}{1}{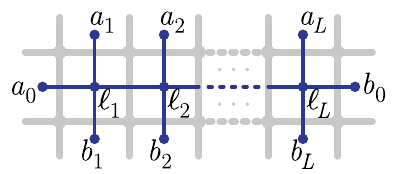}}\label{traintrack_integral_diagram_with_dual}}
with each $\ell_i$ attached to the corresponding loop. Notice that $\x{a_i}{a_{i+1}}\!=\!\x{b_i}{b_{i+1}}\!=\!0$ for $i\!=\!1,\ldots,L\mi1$, corresponding to the requirement that the external particles are massless.   

In terms of these dual coordinates, the Feynman integral (\ref{traintrack_intro}) becomes
\eqL{\traintrack\!\equiv\!\int\!\!d^{4L}\!\vec{\ell}\frac{\prod_{j=0}^L\x{a_j}{b_j}}{\x{\ell_1}{a_0}\Big[\prod_{j=1}^L\x{\ell_{j}}{\ell_{j+1}}\x{\ell_j}{a_j}\x{\ell_j}{b_j}\Big]},\label{dual_space_integrand}}
where $\ell_{L+1}\!\equiv\!b_0$ for notational compactness. The factor in the numerator of (\ref{dual_space_integrand}) has been introduced to ensure that the result is dual-conformally invariant (that is, conformally invariant in dual-momentum $x$-space). As such, the integral should depend exclusively on dual-conformal cross-ratios
\eq{\u{ab}{cd}\!\equiv\!\frac{\x{a}{b}\x{c}{d}}{\x{a}{c}\x{b}{d}}\,.\label{cross_ratio_notation_defined}}

Following the methods described in \mbox{ref.\ \cite{conformalIntegration}}, we Feynman-parameterize the integral (\ref{dual_space_integrand}) within the embedding formalism one loop at a time to obtain
\eqL{\traintrack\!=\!\int\limits_0^{\infty}\!\!\proj{d^{L}\!\vec{\alpha}}\!d^L\!\vec{\beta}\frac{\prod_{j=0}^L\x{a_j}{b_j}}{\big[\x{R_1}{R_1}\cdots\x{R_L}{R_L}\big]\x{R_L}{b_0}},\label{inital_feynman_rep}}
where
\eq{\fwbox{0pt}{R_0\!\equiv\!\alpha_0(a_0),\quad R_k\!\equiv\!(R_{k-1})\pl\alpha_k(a_k)\pl\beta_k(b_k),}}
and $\proj{d^{L}\!\vec{\alpha}}$ represents the projective integration measure over the $L\pl1$ Feynman parameters $\{\alpha_0,\dots, \alpha_L\}$. Note that the $\vec{\beta}$ integration is not projective here (but could be projectivized using the Cheng-Wu theorem \cite{Cheng:1987ga,Smirnov:2006ry}). Rescaling these parameters,
\eq{\fwbox{0pt}{\alpha_k\!\mapsto\!\alpha_k\frac{\x{a_0}{b_k}}{\x{a_k}{b_k}},\quad\beta_k\!\mapsto\!\beta_k\frac{\x{a_0}{a_k}}{\x{a_k}{b_k}},}}
and defining
\eq{\fwbox{0pt}{f_k\!\equiv\!\frac{1}{2}\x{R_k}{R_k}\frac{\x{a_k}{b_k}}{\x{a_0}{a_k}\x{a_0}{b_k}},\quad g_L\!\equiv\!\frac{\x{R_L}{b_0}}{\x{a_0}{b_0}},}}
results in the manifestly dual-conformally invariant expression:
\eq{\traintrack\!=\!\int\limits_0^{\infty}\!\!\proj{d^{L}\!\vec{\alpha}}d^L\!\vec{\beta}\frac{1}{\big(f_1\cdots f_L\big)g_L}\,,\label{dci_parametric_rep}}
where
\vspace{52pt}\eq{\fwbox{0pt}{\hspace{1pt}\,\begin{array}{@{}l@{}l@{}}~\\[-60pt]f_k\!\equiv&\u{a_0a_{k-1}}{a_kb_{k-1}}\hspace{-1pt}\u{a_{k-1}b_k}{b_{k-1}a_0}\hspace{-1pt}\u{a_kb_k}{a_{k-1}b_{k-1}}f_{k-1}\\&\displaystyle\pl\alpha_0(\alpha_k\pl\beta_k)\pl\alpha_k\beta_k\,\pl\!\sum_{j=1}^{k-1}\Big[\alpha_j\alpha_k\u{b_ja_0}{a_ja_k}\\[-2pt]&\displaystyle\hspace{0pt}\pl\alpha_j\beta_k\u{b_ja_0}{a_jb_k}\pl\alpha_k\beta_j\u{a_0a_j}{a_kb_j}\pl\beta_j\beta_k\u{a_0a_j}{b_kb_j}\hspace{-1pt}\Big]\!,\\[-2pt]g_L\!\equiv&\displaystyle\,\alpha_0\,\pl\!\sum_{j=1}^{L}\Big[\alpha_j\u{b_ja_0}{a_jb_0}\pl\beta_j\u{a_0a_j}{b_0b_j}\Big].\\[5pt]~\end{array}}\vspace{-14pt}\label{denominator_factors_defined}}
For the remainder of this work, we will de-projectivize the (otherwise projective) $\vec{\alpha}$ integrations by setting $\alpha_0\!\to\!1$. The form derived above can be easily seen to match the representation of \mbox{ref.\ \cite{Bourjaily:2017bsb}} for $L\!=\!2$ exactly.

Although the collection of cross-ratios arising in (\ref{denominator_factors_defined}) are {\it multiplicatively} independent, the careful reader will note that their number exceeds that of algebraically independent cross-ratios---that is, the dimension of the space of dual-conformal configurations in $x$-space. In this case, the integral $\traintrack$ should depend on $(6L\mi5)$ dual-conformal degrees of freedom. Eliminating these redundancies can be achieved in an elegant way by going to (lower-dimensional) configurations of momentum-twistor space as described in \mbox{ref.\ \cite{Bourjaily:2018aeq}}. 

\vspace{-10pt}\section{Non-Polylogarithmicity}\vspace{-14pt}

The fact that (\ref{dci_parametric_rep}) has $(L\pl1)$ factors in its denominator immediately implies that it has residues of codimension (at least) $(L\pl1)$---signaling at least this degree of `polylogarithmicity'. To see that no further residues exist (without restricting kinematics), it suffices to take the codimension $(L\pl1)$ residue
\eqL{\Res_{\substack{\{f_i=0\}\\g_L=0}}\left(\frac{d^L\!\vec{\beta}\,d\alpha_L}{(f_1\cdots f_L)g_L}\right)=\frac{1}{\sqrt{Q(\alpha_1,\ldots,\alpha_{L-1})}}\label{canonica_residue}}
and observe that $Q$ is generically an irreducible quartic in $\alpha_{L-1}$, and of strictly higher degree in all of the other parameters. Transforming this quartic $Q$ (in $\alpha_{L-1}$) into its \weier representation, we have
\eqL{\frac{1}{\sqrt{Q(\alpha_1,\ldots,\alpha_{L-1})}}\mapsto\frac{1}{\sqrt{4x^3\mi xg_2(\vec{z})\mi g_3(\vec{z})}},}
where $\vec{z}$ denotes the remaining $z_i\!\equiv\!\alpha_i$ for $i\!=\!1,\ldots,L\mi2$. This shows that the residue (\ref{canonica_residue}) results in an integral over an elliptically fibered algebraic variety of dimension $(L\mi1)$,
\eqL{\traintrack=\int\!\!\!\frac{dx\,d^{L\mi2}\vec{z}}{\sqrt{4 x^3\mi g_2(\vec{z})x\mi g_3(\vec{z})}}G'(x,\vec{z}),}
where $G'$ should be some combination of weight-$(L\pl1)$ hyperlogarithms---which we expect will depend in no simple way on the space parameterized by $x,\vec{z}$.

\vspace{-10pt}\section{Geometry of the Elliptic Fibration}\vspace{-14pt}

We have seen that the traintrack integral (\ref{traintrack_intro}) generally involves integration over a space defined by the \weier equation
\eq{y^2=4x^3\mi g_2(\vec{z})x\mi g_3(\vec{z}),\quad x,y\!\in\!\mathbb{C}\,,\label{wS}}
where the complex numbers $\vec{z}$ can be seen as affine coordinates on $\mathbb{P}^{L-2}$, so that the geometry of the space $S$ defined by (\ref{wS}) is that of a complex algebraic variety elliptically fibered over $\mathbb{P}^{L-2}$.

Let us first consider the three-loop instance of (\ref{wS}); we would like to show that the surface $S$ is in fact a K3. For this case, $\vec{z}$ in (\ref{wS}) consists of a single variable, $z$; $g_2(z)$ and $g_3(z)$ are degree 8 and 12 in $z$, respectively; and the variety $S$ is a complex algebraic surface. We can realize this surface as a projective variety by assigning appropriate weights to $x,y,z$, together with an extra homogenizing variable $s$. That is, we may substitute $(x,y,z)\!\mapsto\!(x/s^4, y/s^6, z/s)$ into \eqref{wS} and multiply by an overall factor of $s^{12}$ to make the equation a polynomial. By assigning the degrees of $(x,y,z,s)$ to be \mbox{$[4\!:\!6\!:\!1\!:\!1]$}, the \weier equation \eqref{wS} becomes a {\it homogeneous} degree 12 polynomial in these four variables; thus, we realize the surface $S$ as a homogeneous degree 12 hypersurface in weighted projective space $\mathbb{P}^3_{[1:1:4:6]}$. This is actually one of the K3 surfaces in the classification of refs.\ \cite{Skarke:1996hq,Kreuzer:1998vb} (cf.\ also refs.\ \cite{cydb1,cydb2}) and it is a smooth, compact K3 surface of Picard number two.

More specific features of this K3 surface can be described, but turn out to depend on the kinematic cross-ratios. Although we leave a more thorough study of how the geometry of the K3 varies with the kinematics to future work, let us briefly study how this would be described in the case of some particular point in the space of kinematics. For the sake of illustration, we take the kinematic point corresponding to the $x$-coordinates defined by the function {\tt referenceKinematics[12]} in the package associated with \mbox{ref.\ \cite{Bourjaily:2015jna}} (see also \mbox{refs.\ \cite{Bourjaily:2012gy,Bourjaily:2013mma}}).

First, we check the properties of the \weier discriminant $\Delta(z)$ and the associated $j$-invariant:%
\eq{\fwbox{0pt}{j(z)\equiv \frac{g_2(z)^3}{\Delta(z)},\qquad\Delta(z)\equiv g_2(z)^3\mi 27 g_3(z)^2,}\label{jdelta}}
where $\Delta(z)$ is, in this case, of degree 24. To check the singularities of the fibre, we need to find the roots $z^*_{i=1,\ldots,s}$ of $\Delta(z)$,
and check the orders of $g_2,g_3,\Delta$ at these roots. 
Recall that the order of a function $f(z)$ at $z^*$ is defined as
\eq{\fwbox{0pt}{\Ord_{z^*}\!\big(f(z)\big)\equiv\min_{n\in\mathbb{Z}_{\geq0}}\left\{n:\left.\frac{d^n}{dz^n}f(z)\right|_{z^*} \neq 0\right\};}}
that is, it marks the smallest term in the Laurent expansion of $f(z)$ around $z^*$. For polynomials, this is simply the ordinary order of a root.
We will use the standard notation 
\eq{[n_1^{a_1}, n_2^{a_2}, \ldots, n_s^{a_s}]\label{frame}}
to summarize that there are $a_1$ roots which are of order $n_1$, etc. 
In the context of elliptic fibrations \cite{schuett}, the order is sometimes referred to as the {\it valuation} of $f$.

We also check the order of vanishing of $g_2(z)$ and $g_3(z)$ at each of the roots $z^*_i$ of $\Delta$. We find that $g_{2}(z)$ and $g_{3}(z)$ are both non-vanishing at {\it all} roots $z^*_i$. This means that, according to the Kodaira classification \cite{schuett,kodaira1,kodaira2,kodaira3} of elliptic fibrations, {\it all} singular fibres are of type $I_n$ and the list \eqref{frame} for $\Delta$ is called the Frame shape or cusp-widths \cite{schuett,He:2012jn}.
The fact that all fibres are type $I_n$ means that the elliptic surface is {\it semi-stable}, a member of the class that is perhaps the most well-studied.

Finally, we check the $j$-invariant: it is explicitly a rational function in $z$ and can thus be seen as a map from $\mathbb{P}^1$ onto $\mathbb{P}^1$. It is ramified at \mbox{$\{0,1,\infty\}\!\in\!\mathbb{P}^1$}, meaning that at the preimages of those points the derivative of $j(z)$ could also vanish. In fact, all $8$ preimages of $0$ have ramification index $3$---that is the order of vanishing is 3 there; this is due to the $g_2(z)^3$ in the numerator of (\ref{jdelta}). All 12 preimages of 1 have ramification index 2. Finally, the preimages of $\infty$ (including the double zero at $\infty$ itself) are the roots $z_*$ of $\Delta$. In summary, $j(z)$ is a rational surjective map from $\mathbb{P}^1$ onto $\mathbb{P}^1$ of degree 24; reaffirming that, in addition to being semi-stable and elliptic, $S$ is a K3 surface \cite{schuett}. The Euler number is the degree of $\Delta$ (and the sum of the Frame shape), which is here 24, as is the case for K3. The (ramification indices of the) preimages of 0 can be denoted as $3^8$, those of 1 as $2^{12}$, while those of $\infty$ are summarized by the Frame shape, a partition of 24 which depends on the choice of kinematics. For the kinematic point mentioned above, we find the Frame shape is $[1^8,2^8]$; we have also found kinematics for which the Frame shape is \mbox{$[1^{16},2^4]$}.

There have been various classifications of semi-stable elliptic K3 surfaces (cf.\ refs.\ \cite{shioda,MP,MP2}, especially Props.~3.1--3.7 of ref.\ \cite{MP} for allowed Frame shapes) and our above example of $[1^8,2^8]$ is present in these classifications.
We remark that for certain large classes of semi-stable elliptic fibrations (e.g., when a Riemann-Hurwitz condition (cf.~Theorem 2.3 of ref.\ \cite{BM}) is further obeyed), $j(z)$ has a very beautiful property: it is Belyi  \cite{schuett,He:2012jn}. This means that $j(z)$ is ramified at {\it only} the three points \mbox{$\{0,1,\infty\}\!\in\!\mathbb{P}^1$}. This happens, for instance, in the situation of {\it extremal} semi-stable elliptic K3 surfaces which have six preimages of $\infty$, so that the Frame shape is a 6-partition of 24. For such cases, one can associate {\it dessins d'enfant} to $S$, which have important number-theoretic properties. It would therefore be very interesting to find kinematic points which yield a $\Delta$ with six distinct roots.

\vspace{-10pt}\section{Calabi-Yaus at Higher Loops}\vspace{-10pt}
Let us consider one further example in some detail: the case of $L\!=\!4$. In this case, the \weier equation (\ref{wS}) defines a complex threefold as an elliptic fibration over a base parameterized by $z_1,z_2\!\in\!\mathbb{C}$. Explicitly, we may gather the terms $g_{2,3}(z_1,z_2)$ in \eqref{wS} by their degrees according to,
\eq{\begin{split}g_2(\vec{z})&\equiv\sum_{k=-4}^4z_2^{4-k}g_2^{(12+k)}(z_1),\\g_3(\vec{z})&\equiv\sum_{k=-6}^{6}z_2^{6-k}g_3^{(18+k)}(z_1),\end{split}\label{wX}}
where $g_{i=2,3}^{(k)}$ are polynomials of degree $k$ in $z_1$ only.
Suppose we projectivized (\ref{wS}) by embedding it into  $\mathbb{P}^4_{[8:12:1:1:1]}[x,y,z_1,z_2,s]$; we would end up with a homogeneous degree 24 hypersurface.
The sum of weights, however, equals 23 and thus explicitly we seemingly violate the Calabi-Yau condition.
This need not worry us, however, since it merely means that the variety $S$ defined by \eqref{wS} with $g_i$ as in \eqref{wX}, if it were Calabi-Yau, 
is not realizable as a hypersurface in weighted $\mathbb{P}^4$.
Nevertheless, we have explicitly checked using \cite{M2}, for a general choice of coefficients in \eqref{wX}, that the surface $S$ is irreducible, smooth, and complex dimension 3, together with the canonical sheaf $K_S\!\simeq\!\bigwedge^3 T^*_S$ being the trivial line-bundle $\mathcal{O}_S$; therefore, $S$ is a Calabi-Yau threefold. Since there is a classification of elliptic Calabi-Yau threefolds \cite{Morrison:1996pp}, it would be interesting to identify our particular $S$.

We expect this behavior to continue to higher loops, with the elliptically fibered variety described by (\ref{wS}) always being a Calabi-Yau $(L\mi1)$-fold. We leave the proof of this, together with a more thorough investigation of the specific varieties along this sequence, to future work. 

\vspace{-10pt}\section{Traintracks As Amplitudes in \texorpdfstring{$\mathcal{N}\!=\!4$}{N=4} SYM}\vspace{-10pt}

Let us now demonstrate the claim made in the introduction that the traintrack integral (\ref{traintrack_intro}), in addition to contributing to massless $\phi^4$ theory, is an unavoidable contribution to amplitudes in planar $\mathcal{N}\!=\!4$ SYM theory. This is a simple consequence of the fact that there exists a component amplitude of planar $\mathcal{N}\!=\!4$ SYM theory for which (\ref{traintrack_intro}) represents the entire (leading) contribution! That is, we want to show that the $L$-loop contribution to the component amplitude
\vspace{-2.5pt}\eqL{\mathcal{A}\big(\overbrace{{\color{hblue}\phi_{12}},{\color{hblue}\ldots},{\color{hblue}\phi_{12}}}^{L+1},{\color{hred}\phi_{13}},{\color{hred}\phi_{13}},\overbrace{{\color{hblue}\phi_{34}},{\color{hblue}\ldots},{\color{hblue}\phi_{34}}}^{L+1},{\color{hred}\phi_{24}},{\color{hred}\phi_{24}}\big)\label{sym_component}\vspace{-2pt}}
is computed by the single scalar Feynman integral
\eq{\hspace{-0pt}\fwbox{0pt}{\fig{-22.6pt}{1}{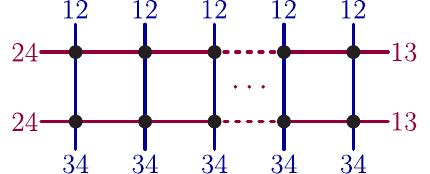}\,.}\label{fishnet_diagram}}
This integral is the same as $\traintrack$ (upon dividing by the numerator introduced in (\ref{dual_space_integrand})). To be clear, (\ref{sym_component}) denotes the bosonic component
\eq{\begin{split}\int\!\!&\big(d\tilde{\eta}_1^1\!\cdots\! d\tilde{\eta}^1_{L+3}\big)\big(d\tilde{\eta}_{n-1}^2\!\cdots\! d\tilde{\eta}^2_{L+1}\big)\\
&\big(d\tilde{\eta}_{L+2}^3\!\cdots\! d\tilde{\eta}^3_{n-2}\big)\big(d\tilde{\eta}_{L+4}^4\!\cdots\! d\tilde{\eta}^4_{n}\big)\mathcal{A}^{(k=L+1),L}_{n=2L+6}\end{split}}
of the $(2L\pl6)$-point N$^{L+1}$MHV amplitude, where $\tilde{\eta}$ are the Gra\ss mann variables of Nair's $\mathcal{N}\!=\!4$ on-shell superfield \cite{Nair:1988bq}. This component vanishes below $L$ loops. 

The claim above has already been noted for \mbox{$L\!=\!2$} loops by the authors of \mbox{ref.\ \cite{CaronHuot:2012ab}}. The general case follows as a simple consequence of the relationship between planar $\mathcal{N}\!=\!4$ SYM theory and an integrable fishnet theory \mbox{\cite{Gurdogan:2015csr,Sieg:2016vap,Grabner:2017pgm}}. 

In this fishnet theory, there exists only a single single-trace interaction vertex, $\mathrm{Tr}\big({\color{hblue}\phi^{12}}{\color{hred}\phi^{13}}{\color{hblue}\phi^{34}}{\color{hred}\phi^{24}}\big)$. This theory is not unitary, but it is (dual) conformal---and indeed, integrable---in the planar limit. It is easy to appreciate that very few Feynman diagrams contribute to the fishnet theory at any loop order; and likewise easy to see that (\ref{fishnet_diagram}) is the only one which contributes to the amplitude for this configuration of fields in the planar limit. 

The fact that (\ref{fishnet_diagram}) is also the (entire) amplitude in planar $\mathcal{N}\!=\!4$ SYM theory follows from the way in which the fishnet theory is obtained as a certain double-scaling limit of the $\gamma_i$-deformation of $\mathcal{N}\!=\!4$ SYM theory \cite{Frolov:2005dj}. This deformation can be formulated by replacing every product in the action by a Moyal-like $\star$-product \cite{Lunin:2005jy}, which introduces a phase that depends on the $SU(4)_R$ charges of the respective fields. There is a theorem \cite{Filk:1996dm} that every planar Feynman diagram in the deformed theory is given by its value in the undeformed theory times the phase of the $\star$-product of its external field in cyclic order. As a corollary, the same relation holds for planar scattering amplitudes \cite{Khoze:2005nd}.\footnote{However, this theorem cannot be applied to form factors or correlation functions of operators that carry nontrivial $R$-charge \cite{Fokken:2013mza}.} This concludes the proof that the component amplitude (\ref{sym_component}) is (up to some constant) given by the same integral as in the fishnet theory---namely, (\ref{fishnet_diagram}).

Finally, let us note the stark contrast between the rich structure we uncover here in planar scattering amplitudes in $\mathcal{N}\!=\!4$ SYM theory and the planar spectrum of scaling dimensions of composite operators in this theory, which via integrability is conjectured to be given by multiple zeta values at all loop orders \cite{Leurent:2013mr}.

\vspace{-10pt}\section{Conclusions}\vspace{-10pt}

We have defined a class of `traintrack' integrals \eqref{traintrack_intro} that increase in complexity with increasing loop order, from dilogarithms at one loop to elliptic behavior at two loops to a K3 at three loops and a Calabi-Yau threefold at four loops. The curves defined by this sequence are elliptically fibered, and we have conjectured that they are Calabi-Yau to any order. These diagrams appear in massless $\phi^4$ theory and are the sole contribution to a specific component amplitude in planar ${\cal N} = 4$ SYM theory. More generally, we expect them to contribute to a wide range of quantum field theories beyond these examples.

Working with these integrals will demand the development of new tools beyond those currently available. The symbol and coaction of polylogarithmic functions have proven extremely useful, and similar tools have recently been developed for elliptic polylogarithms \cite{brown2011multiple,Broedel:2018iwv}. Such tools would seem to be much more difficult to develop for K3 and higher functions. Still, it would be interesting to see if these functions can be understood one day on the same level as their polylogarithmic cousins.

It would be intriguing to investigate the behavior of these functions to all orders. Ladder integrals have been understood to all orders for quite some time, and have even been resummed in the coupling \cite{Usyukina:1992jd,Usyukina:1993ch,Broadhurst:1993ib,Broadhurst:2010ds,Drummond:2012bg}. Traintracks are more challenging, as their kinematic dependence changes at each order. We could hope at least to understand the curves they give rise to---whether these curves truly are Calabi-Yau to all orders, and what sort of sequence of curves they generate. Our traintracks might then serve as a safe path through the vast Calabi-Yau wilderness, our Virgil as we move deeper into these geometries.

\vspace{6pt}\acknowledgments
We thank  David Broadhurst, Johannes Br\"{o}del, Freddy Cachazo, Dan Grayson, Erik Panzer, Richard Thomas, Pierre Vanhove, and Raimundas Vidunas for stimulating discussions, and we are grateful to Marcus Spradlin and Pierre Vanhove for thoughtful comments on early drafts of this Letter. 
This work was supported in part by the Danish Independent Research Fund under grant number DFF-4002-00037 (MW); Danish National Research Foundation under grant number DNRF91, a grant from the Villum Fonden and a Starting Grant \mbox{(No.\ 757978)} from the European Research Council (JLB,AJM,MvH,MW). YHH would like to thank the Science and Technology Facilities Council, UK, for grant ST/J00037X/1, the Chinese Ministry of Education, for a Chang-Jiang Chair Professorship at NanKai University as well as the City of Tian-Jin for a Qian-Ren Scholarship, and Merton College, Oxford, for her enduring support.


\vspace{-10pt}\providecommand{\href}[2]{#2}\begingroup\raggedright\endgroup

\end{document}